\documentclass[12pt]{iopart}

\usepackage{graphicx}

\begin{document}

\title[Simulation of interaction Hamiltonians by quantum feedback]{Simulation of interaction Hamiltonians\\
by quantum feedback:\\
a comment on the dynamics of information exchange between
coupled systems}


\author{Holger F. Hofmann}
\address{
Graduate School of Advanced Sciences of Matter, Hiroshima University,
Kagamiyama 1-3-1, Higashi Hiroshima 739-8530, Japan}

\begin{abstract}
Since quantum feedback is based on classically accessible measurement
results, it can provide fundamental insights into the dynamics of
quantum systems by making available classical information on the
evolution of system properties and on the conditional
forces acting on the system.
In this paper, the feedback-induced interaction dynamics
between a pair of quantum systems is analyzed. It is pointed out that
any interaction Hamiltonian can be simulated by local feedback if
the levels of decoherence are sufficiently high. The boundary
between genuine entanglement generating quantum interactions and
non-entangling classical interactions is identified and the
nature of the information exchange between two quantum systems during
an interaction is discussed.
\end{abstract}


\ead{h.hofmann@osa.org}

\maketitle

\section{Introduction}
In recent years, there has been rapid progress in the analysis
of quantum systems at a fundamental level based on precise
measurements and optimized control. Experimentally, optical
systems have provided a fertile testing ground for this new
understanding of quantum effects due to the availability of
highly coherent lasers and sensitive detectors.
In the light of these new technologies, the discussion about
quantum measurement has obtained a practical relevance that
sometimes challenges the seemingly well-defined notions of
quantum states and Hamiltonians conveyed by typical introductions
to quantum mechanics.

A particularly instructive example is the
development of quantum feedback theory, which originated from
a formal analysis of open system dynamics that was largely motivated
by the intention to identify the proper pure state description in
the presence of noise, but resulted instead in a new formulation of
conditional quantum dynamics that actually highlights the inadequacy
of a measurement independent definition of pure states in the
dynamics of open systems \cite{Dal92,Wis93a,Car93,Wis93b,Wis94}.
By emphasizing the importance of classically available measurement
information, the theory of quantum feedback then permitted the
identification of interesting parallels between the classical notion
of control and its quantum mechanical equivalent \cite{Doh00}, a result
that may be of significant practical use.

My own interest in quantum feedback originated from studies of
quantum noise in lasers, where the classical description of light
is surprisingly successful even when the intensities studied are
so small that individual photons could be resolved. This observation
can be explained in detail by quantum measurement theory: in the presence
of a strong laser field, the main effect of the spontaneous emission from
a single atom is its interference with the laser light, not the energy
contributed by the single atom. It is thus reasonable to analyze the
dynamics of single atom emission in terms of homodyne detection of the
electromagnetic field. Interestingly, quantum dynamics then corresponds
closely to classical electrodynamics, and the effects of quantum noise
and of classical dipole radiation can be identified in the measurement
statistics. By using feedback, it is therefore possible to eliminate the
quantum noise effects and to reduce the effect of spontaneous emission
to a quantum nondemolition measurement of the atomic dipole \cite{Hof98a,Hof98b}. This method may have interesting applications to
the stabilization of atomic quantum states, and it has been pointed
out that, in principle,
this kind of quantum feedback can indeed stabilize almost any quantum
state of a two level atom \cite{Wan01}.

Quantum feedback can thus restore some of the classical concepts of
dynamics that seem to be lost in the transition from classical systems
to quantum systems. In particular, the measurement information used in
feedback can always be interpreted as a minimal back-action
measurement of a system variable and a conditional unitary evolution
of the system \cite{Wis95}. A feedback setup therefore keeps track of some
of the observable properties of the system, in addition to the information
available on the conditional forces responsible for the deterministic
evolution of the system \cite{Wis96}. It may be worthwhile to
persue this line of thought a bit further in order to gain a better
understanding of the relationship between the unobservable closed
system dynamics and the observable (and therefore controllable)
dynamics of quantum feedback. Specifically, this approach could shed some
light on the quantum mechanical interaction dynamics between two
coupled systems by providing a description of interaction that combines
the classical notion of deterministic conditional forces with the
quantum notion of the interaction Hamiltonian.

In the following, I will therefore describe systems where the interaction
is realized entirely by quantum feedback. According to standard quantum
feedback theory, the dynamics is then described by an effective
interaction Hamiltonian and a (seemingly separate) superoperator describing
the decoherence associated with the measurement interaction.
The feedback system can thus simulate a genuine quantum interaction.
By itself, the effective Hamiltonian would entangle the interacting system,
but in the context of the feedback setup, the quantum state must remain
separable because the quantum operations on the individual systems
are in fact local. In the terminology of quantum information theory
\cite{Nie00}, the interaction has been realized by local
operations and classical communication
between the systems, whereas the Hamiltonian interaction itself corresponds
to quantum communication between the systems.

The result of this analysis shows that the decoherence term is not
really separable from the Hamiltonian dynamics, but describes the quantum noise required to reduce the exchange of quantum information between
the systems to zero. The transition from a purely classical exchange
of traceable information to the more intimate entanglement
generating quantum interaction
is therefore a quantitative one based on the precise relation between the
noise levels and the coupling constants representing the forces acting
between the systems.
This observation is consistent with recent results on the
robustness of quantum gate operations against noise
\cite{Har03,Mon03,Ban04,Hof04a,Hof04b} and may therefore have
interesting implications for the evaluation of experimental quantum
devices.
The quantum feedback analysis of interactions between quantum systems thus
reveals an interesting connection between basic concepts of quantum
information and physical interactions. Moreover, the qualitative
correspondence between the feedback dynamics and the quantum interaction
described by the Hamiltonian indicates that the interpretation of
interactions in terms of conditional forces acting between systems might
have its applications even in the context of genuine quantum interactions.

The rest of the paper is organized as follows.
In section \ref{sec:qf}, the possibility of modifying the
Hamiltonian dynamics by quantum feedback is reviewed.
In section \ref{sec:int}, these results are applied to a pair
of non-interacting systems to generate an effective interaction
Hamiltonian. In section \ref{sec:limits}, the results are applied to
the analysis of noisy interactions to derive a criterion for the
separability of the interaction dynamics. In section
\ref{sec:squeeze}, the results are illustrated for the case of
a feedback induced interaction between a pair of optical cavity modes.
It is shown that the feedback analysis provides exact uncertainty
limits for the separability of the two mode squeezing interaction.
In section \ref{sec:discuss}, the implications of separability
for the information exchange between interacting quantum systems
is discussed. The results are summarized in \ref{sec:concl}.

\section{Quantum feedback and effective Hamiltonians}
\label{sec:qf}

First, it may be useful to review some of the central results of
continuous quantum feedback theory \cite{Wis94} in terms of the relation
between measurement information and conditional dynamics.
For this purpose, let us consider the dynamics of a quantum system
coupled to the environment in such a way that some observable
property $\hat{X}$ causes the emission of a corresponding signal
in a quantum fluctuating field propagating away from the system
to a detector setup. The effects of the environment on the
system state then causes dephasing between eigenstates of $\hat{X}$.
If this is the only relevant dynamics
of the system, the evolution of the density matrix $\hat{\rho}$
can be written as
\begin{equation}
\label{eq:qnd}
\frac{d}{dt} \hat{\rho} = \gamma \left(\hat{X} \hat{\rho} \hat{X}
- \frac{1}{2} \hat{X}^2 \hat{\rho} - \frac{1}{2} \hat{\rho} \hat{X}^2
 \right),
\end{equation}
where $\gamma$ is the coupling rate determining the strength of the
interaction between the system variable $\hat{X}$ and the signal
field.
At the detector, the emitted signal can then be measured with a
resolution limited by the quantum fluctuations of the field \cite{Car93}.
If the signal is integrated over a finite time interval $\Delta t$,
this resolution is given by
\begin{equation}
\label{eq:resolution}
\frac{1}{\delta \! X^2} = 4 \gamma \Delta t,
\end{equation}
where $\delta \! X$ is the expected error in the measurement result
$X_m$ of the observable $\hat{X}$ obtained during the time interval
$\Delta t$.

The measurement result $X_m$ is now a classical record of the quantum
property $\hat{X}$. In quantum feedback, this record is used to
condition the quantum dynamics of the system \cite{Wis94}.
In the case of linear feedback, the feedback can be described by a
Hamiltonian of the form
\begin{equation}
\label{eq:Hfb}
\hat{H}_{\mbox{feedback}} (X_m) = - 2 \hbar \gamma X_m \hat{Y}.
\end{equation}
If the time delay between the emission of the signal and the application
of the feedback can be neglected, the effective dynamics of the system
can be obtained by applying the operators representing the measurement
and the feedback for the time interval $\Delta t$ to both sides of the
density matrix. The feedback dynamics of the density matrix can then
be determined by averaging over all possible measurement results.
Due to this averaging procedure, only quadratic terms in
$\hat{X}$ and $\hat{Y}$ contribute, and the result reads
\begin{eqnarray}
\label{eq:fb1}
\frac{d}{dt} \hat{\rho} &=& \gamma \left(\hat{X} \hat{\rho} \hat{X}
- \frac{1}{2} \hat{X}^2 \hat{\rho} - \frac{1}{2} \hat{\rho} \hat{X}^2
\right)
\nonumber \\ &+&
\gamma \left(\hat{Y} \hat{\rho} \hat{Y}
- \frac{1}{2} \hat{Y}^2 \hat{\rho} - \frac{1}{2} \hat{\rho} \hat{Y}^2
\right)
\nonumber \\ &+&
\gamma \left(i \hat{Y} \hat{\rho} \hat{X} - i \hat{X} \hat{\rho} \hat{Y}
+ i \hat{Y} \hat{X} \hat{\rho} - i \hat{\rho} \hat{X} \hat{Y}
\right).
\end{eqnarray}
In this representation, the first term represents the original dephasing
in $\hat{X}$ caused by the emission of the signal, the second term
represents the dephasing in $\hat{Y}$ caused by the quantum noise in
the feedback, and the third term represents the effects of the correlation
between the measurement results obtained for $\hat{X}$ and the feedback
defined by $\hat{Y}$. Note that the third term is not symmetric in
$\hat{X}$ and $\hat{Y}$ due to the temporal sequence of the feedback.
Where $\hat{X}$ and $\hat{Y}$ are applied to the same side of $\hat{\rho}$,
the measurement term $\hat{X}$ is always applied before the feedback
term $\hat{Y}$. As will be shown below, this physically motivated sequence
establishes a necessary correlation between the unitary evolution and the
decoherence of the system.

The dynamics given by equation (\ref{eq:fb1}) above can be transformed
to a more conventional form by ordering the operator products
according to the position of $\hat{\rho}$,
\begin{eqnarray}
\label{eq:fb2}
\frac{d}{dt} \hat{\rho} &=& \gamma
\Big((\hat{X}+i\hat{Y}) \hat{\rho} (\hat{X}-i\hat{Y})
\nonumber \\ &&
- \frac{1}{2} (\hat{X}^2 - i 2 \hat{Y}\hat{X} + \hat{Y}^2)
\hat{\rho}  - \frac{1}{2} \hat{\rho}
(\hat{X}^2 + i 2 \hat{X}\hat{Y} + \hat{Y}^2) \Big).
\end{eqnarray}
Here, the effects of measurement and feedback have been combined in
such a way that their different physical origin is not recognizable
anymore. In fact, the dynamics can now be summarized using only two
operators, one to describe decoherence, and one to describe the unitary
evolution of the quantum state. The dynamics then read
\begin{equation}
\label{eq:fb3}
\frac{d}{dt} \hat{\rho} = \gamma
\left(\hat{c} \hat{\rho} \hat{c}^\dagger
- \frac{1}{2} \hat{c}^\dagger \hat{c} \hat{\rho}
- \frac{1}{2} \hat{\rho} \hat{c}^\dagger \hat{c}
\right)
- \frac{i}{\hbar} [ \hat{H}_{\mbox{eff.}}, \hat{\rho} ],
\end{equation}
where the decoherence operator is $\hat{c}=\hat{X}+i\hat{Y}$ and the
effective Hamiltonian $\hat{H}_{\mbox{eff.}}$ is given by
\begin{equation}
\label{eq:Heff}
\hat{H}_{\mbox{eff.}} = - \frac{\hbar \gamma}{2}
\left( \hat{X}\hat{Y}+\hat{Y}\hat{X} \right).
\end{equation}
It is interesting to compare this effective Hamiltonian with the
actual feedback operator of equation (\ref{eq:Hfb}). The essential
difference is that the measurement value $X_m$ has now been replaced
with the operator $\hat{X}$. While this replacement appears to be
a rather intuitive result since $X_m$ is the measurement result
of the observable property represented by $\hat{X}$, it is important
to recognize that the replacement of a (generally continuous) real
number with a hermitian operator completely changes the dynamics
described by the Hamiltonian. In particular, the effective Hamiltonian
is symmetric in $\hat{X}$ and $\hat{Y}$, indicating that the same
Hamiltonian could be obtained from a measurement of $\hat{Y}$ followed
by a feedback in $\hat{X}$. The difference between the two scenarios
only appears in the exchanged roles of $\hat{c}$ and $\hat{c}^\dagger$.

Equation (\ref{eq:fb3}) now gives the feedback dynamics of a system
emitting a signal dependent on only a single observable of the system.
The dynamics are therefore based on the case of an ideal quantum
non-demolition measurement, as described by equation (\ref{eq:qnd}).
Nevertheless the resulting decoherence operators $\hat{c}$ in equation 
(\ref{eq:fb3}) have the form of annihilation operators, and the 
decoherence dynamics after feedback appears similar to typical photon 
emission dynamics. Interestingly, this observation has a simple 
classical explanation.
In classical electrodynamics, radiation losses can be described by
the back-action of the emitted electric field on the oscillating dipole
itself. If the emitted dipole field is known, this kind of back-action
can be represented by a feedback Hamiltonian. As reported in
\cite{Hof98a,Hof98b}, homodyne detection of the emitted radiation can be
used to identify and to compensate this back-action by applying
a feedback Hamiltonian that cancels the effects of this deterministic
back-action.
It is thus possible to interpret the natural relaxation processes of
a system interacting with the environment in terms of a hypothetical
combination of minimal back-action measurement and quantum feedback,
where the feedback represents the quantum version of classical
back-action \cite{Wis95}.

The analysis given above shows how the classical information flow
in a quantum feedback system can be identified with elements of
the quantum dynamics of the density matrix. In the next section,
we will consider the consequences of these results for interactions
between two separate systems by analyzing the simulation of
interaction Hamiltonians by quantum feedback.

\section{Information dynamics in basic interactions}

\label{sec:int}

\begin{figure}
\begin{center}
\begin{picture}(280,200)
\thicklines
\put(30,90){\makebox(20,20){\Large $A$}}
\put(70,100){\circle{40}}
\put(60,90){\makebox(20,20){\Large $X_A$}}

\put(230,90){\makebox(20,20){\Large $B$}}
\put(210,100){\circle{40}}
\put(200,90){\makebox(20,20){\Large $X_B$}}

\bezier{200}(70,120)(100,160)(140,160)
\bezier{200}(140,160)(180,160)(210,120)
\put(140,160){\line(-3,2){10}}
\put(140,160){\line(-3,-2){10}}
\put(120,170){\makebox(40,20){\Large $F_{A \to B}(X_A)$}}
\bezier{200}(70,80)(100,40)(140,40)
\bezier{200}(140,40)(180,40)(210,80)
\put(140,40){\line(3,2){10}}
\put(140,40){\line(3,-2){10}}
\put(120,10){\makebox(40,20){\Large $F_{B \to A}(X_B)$}}

\end{picture}
\end{center}
\caption{\label{fig1}
Illustration of the information exchange in the classical interaction
between a system $A$ and a system $B$.
The property $X_A$ of system $A$ causes the action of a force
$F_{A \to B}$ in system $B$, and vice versa.}
\end{figure}
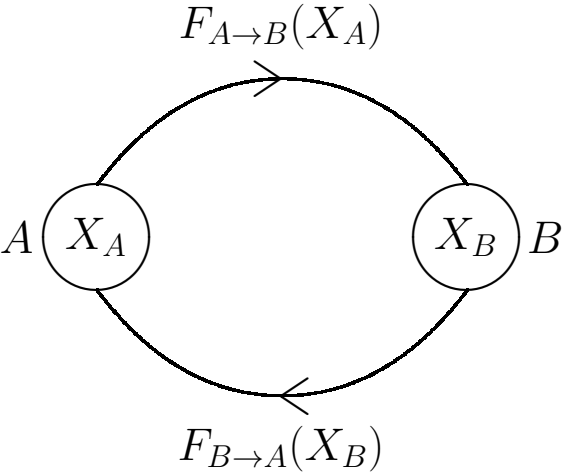

In classical physics, interactions can always be described in
terms of local forces acting on local systems. The fact that
the forces originate from other systems can be taken into
account by simply correlating the specific value of the force
with the value of the corresponding system property. A classical
interaction between a system $A$ and a system $B$ can thus be
described as shown in figure \ref{fig1}: A force $F_{A \to B}$
depending on the value of the property $X_A$ of system $A$
acts locally on system $B$, while a force $F_{B \to A}$
depending on the value of the property $X_B$ of system $B$
acts locally on system $A$. The dynamics of each system can
therefore be treated locally, while the connection between the
systems is established by the transfer of the values of $X_A$
and of $X_B$ from one system to the other. This transfer of
classical variables corresponds to a classical communication
channel and could be realized by a classical feedback line using
precise measurement data from the remote system.

\begin{figure}
\begin{center}
\begin{picture}(280,200)
\thicklines
\put(30,90){\makebox(20,20){\Large $A$}}
\put(70,100){\circle{40}}
\put(60,90){\makebox(20,20){\Large $\hat{X}_A$}}

\put(230,90){\makebox(20,20){\Large $B$}}
\put(210,100){\circle{40}}
\put(200,90){\makebox(20,20){\Large $\hat{X}_B$}}

\bezier{200}(70,120)(100,160)(140,160)
\bezier{200}(140,160)(180,160)(210,120)
\bezier{200}(70,80)(100,40)(140,40)
\bezier{200}(140,40)(180,40)(210,80)
\put(130,90){\makebox(20,20){\LARGE $\hat{H}_{AB}$}}
\put(95,100){\line(1,0){28}}
\put(95,100){\line(1,1){10}}
\put(95,100){\line(1,-1){10}}
\put(185,100){\line(-1,0){28}}
\put(185,100){\line(-1,1){10}}
\put(185,100){\line(-1,-1){10}}
\put(50,10){\makebox(180,20){\Large $\hat{\rho}_{AB}$ not separable}}
\end{picture}
\end{center}
\caption{\label{fig2}
Illustration of the closed-system interaction between two
quantum systems $A$ and $B$. The interaction Hamiltonian $
\hat{H}_{AB}$ acts on the joint state $\hat{\rho}_{AB}$ of both
systems. As a result of the interaction, $\hat{\rho}_{AB}$ is
usually not separable into a product state of $A$ and $B$.
}
\end{figure}
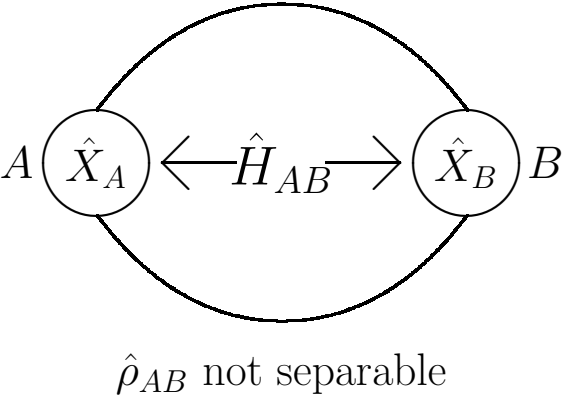

In the quantum case, things are a bit more complicated. As illustrated
by figure \ref{fig2}, the Hamiltonian $\hat{H}_{AB}$ describing the
interaction dynamics necessarily acts on both systems at once. In general,
it is not possible to define the local dynamics, as evidenced by the
possibility of generating non-separable entangled states in the interaction.
Effectively, the systems exchange genuine quantum information, and their
interaction cannot be described in terms of local operations based on
classical communication between the systems. As a result of this
quantum communication, the density matrix $\hat{\rho}_{AB}$ usually
becomes entangled and cannot be represented by products of local density
matrices. However, it is a well-known fact that entangled
states become separable when a given amount of noise is added. Likewise,
quantum interactions can be analyzed in terms of a feedback model
closely corresponding to the classical case shown in figure
\ref{fig1} if the decoherence rate of the interacting systems is
sufficiently high.

\begin{figure}
\begin{center}
\begin{picture}(280,200)
\thicklines
\put(20,90){\makebox(20,20){\Large $A$}}
\put(60,100){\circle{40}}
\put(50,90){\makebox(20,20){\Large $\hat{X}_A$}}

\put(240,90){\makebox(20,20){\Large $B$}}
\put(220,100){\circle{40}}
\put(210,90){\makebox(20,20){\Large $\hat{X}_B$}}

\bezier{200}(150,160)(190,160)(220,120)
\put(220,120){\line(0,1){10.5}}
\put(220,120){\line(-3,1){10}}

\bezier{200}(60,120)(90,160)(125,160)
\put(125,160){\line(-3,2){10}}
\put(125,160){\line(-3,-2){10}}
\put(50,150){\makebox(40,20){\Large $X_{m,A}$}}
\put(135,145){\line(0,1){30}}
\bezier{100}(135,145)(150,145)(150,160)
\bezier{100}(135,175)(150,175)(150,160)
%
%
\bezier{200}(60,80)(90,40)(130,40)
\put(60,80){\line(0,-1){10.5}}
\put(60,80){\line(3,-1){10}}
\put(190,150){\makebox(80,20){\Large $\hat{H}_{A \to B}(X_{m,A})$}}

\bezier{200}(155,40)(190,40)(220,80)
\put(155,40){\line(3,2){10}}
\put(155,40){\line(3,-2){10}}
\put(190,25){\makebox(40,20){\Large $X_{m,B}$}}
\put(145,25){\line(0,1){30}}
\bezier{100}(145,25)(130,25)(130,40)
\bezier{100}(145,55)(130,55)(130,40)

\put(20,25){\makebox(80,20){\Large $\hat{H}_{B \to A}(X_{m,B})$}}

\end{picture}
\end{center}
\caption{\label{fig3}
Illustration of a quantum feedback setup realizing an effective
Hamiltonian of $\hat{H}_{\mbox{eff.}}=2 \hbar \gamma \hat{X}_A\hat{X}_B$.]
The property $\hat{X}_k$ of each system emits an observable signal
$X_{m,k}$ measured at the detectors. This measurement value is then
used to define a feedback Hamiltonian $\hat{H}_{k \to l}$ acting on
the opposite system $l$ ($k,l=A,B$).
}
\end{figure}
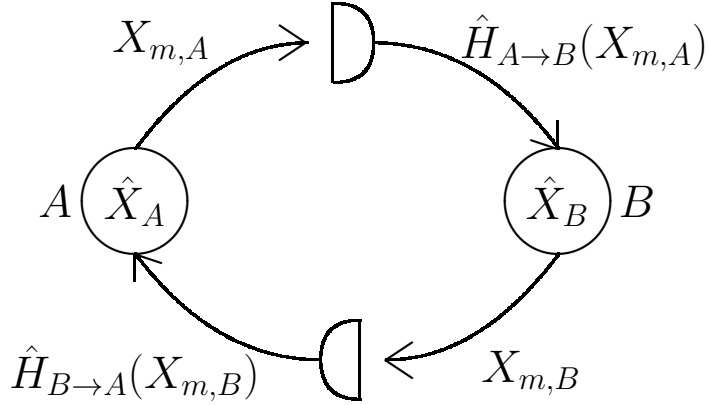

Figure \ref{fig3} shows the quantum feedback setup realizing
an effective Hamiltonian of $2 \hbar \gamma \hat{X}_A \hat{X}_B$ by
correlated local operations on systems $A$ and $B$. In this setup,
the systems are now connected by classical communication lines carrying
information on the measurement outcomes $X_{m,A}$ and
$\hat{X}_{m,B}$ for the system properties $\hat{X}_A$ and $\hat{X}_B$
from one system to the other.
Using the quantum feedback
theory introduced in section \ref{sec:qf} above, it is then possible
to determine the effective interaction dynamics between the systems.
Specifically, the initial decoherence dynamics without feedback
can be written as
\begin{eqnarray}
\label{eq:qndAB}
\frac{d}{dt} \hat{\rho}_{AB} &=& D(\hat{\rho}_{AB}), \hspace{0.5cm} \mbox{with}
\nonumber \\[0.2cm]
D(\hat{\rho}_{AB}) &=&
 \gamma \left(\hat{X}_A \hat{\rho}_{AB} \hat{X}_A
- \frac{1}{2} \hat{X}_A^2 \hat{\rho}_{AB} - \frac{1}{2} \hat{\rho}_{AB}
\hat{X}_A^2 \right)
\nonumber \\ &&
+ \gamma \left(\hat{X}_B \hat{\rho}_{AB} \hat{X}_B
- \frac{1}{2} \hat{X}_B^2 \hat{\rho}_{AB} - \frac{1}{2} \hat{\rho}_{AB}
\hat{X}_B^2 \right).
\end{eqnarray}
where the indices of the operators indicate which system they act on.
This decoherence permits a continuous measurement of $\hat{X}_A$
and $\hat{X}_B$ with a resolution of $4 \gamma \Delta t$ per
time interval $\Delta t$ as given by equation (\ref{eq:resolution}).
The measurement results $X_{m,A}$ and $X_{m.B}$ obtained during each
time interval can then be used to generate a linear feedback acting
on the opposite system with feedback Hamiltonians given by
\begin{eqnarray}
\label{eq:HfbAB}
\hat{H}_{A\to B} (X_{m,A}) &=& - 2 \hbar \gamma X_{m,A} \hat{X}_B
\hspace{0.5cm}\mbox{and}
\nonumber \\
\hat{H}_{B\to A} (X_{m,B}) &=& - 2 \hbar \gamma X_{m,B} \hat{X}_A.
\end{eqnarray}
The feedback dynamics can then be determined most effectively
by using equation (\ref{eq:fb2}), since a number of imaginary
terms generated by the two feedback lines cancel. As a result,
the decoherence term retains its original form, with the
decoherence rate being doubled by the noise in the feedback.
The joint dynamics of the feedback-coupled systems can then be
written as
\begin{equation}
\label{eq:coupled}
\frac{d}{dt} \hat{\rho}_{AB} =
- \frac{i}{\hbar} [ \hat{H}_{\mbox{eff.}}, \hat{\rho}_{AB} ]
+ 2 D(\hat{\rho}_{AB}),
\end{equation}
where the effective interaction Hamiltonian is given by
\begin{equation}
\label{eq:HeffAB}
\hat{H}_{\mbox{eff.}} = 2 \hbar \gamma \hat{X}_A \hat{X}_B.
\end{equation}
The non-local quantum interaction represented by the Hamiltonian
$\hat{H}_{\mbox{eff.}}$ above can thus be simulated by a feedback
setup in which only classical information is exchanged between the
systems. The price to be paid for the replacement of quantum interactions
with a classical signal transfer is given by the decoherence
operator $2 D$.
The qualitative effects of the interaction Hamiltonian
$\hat{H}_{\mbox{eff.}}$ can then be analyzed in terms of an
$\hat{X}_A$-dependent unitary transform acting on system $B$ and a
$\hat{X}_B$-dependent unitary transform acting on $A$.

More complicated interactions can be simulated if measurement information
on other system variables is available in the emitted fields. In principle,
any interaction Hamiltonian can be simulated by decomposing it into
a sum of bilinear terms of the form $2 \hbar \gamma \hat{X}_A \hat{X}_B$,
implementing each term by a separate feedback. Quantum feedback interactions
can therefore be used to implement a wide range of interaction
Hamiltonians between systems that are only connected by classical
communication lines.

\section{Interactions in
a noisy environment:
quantum limits of decoherence rates}

\label{sec:limits}

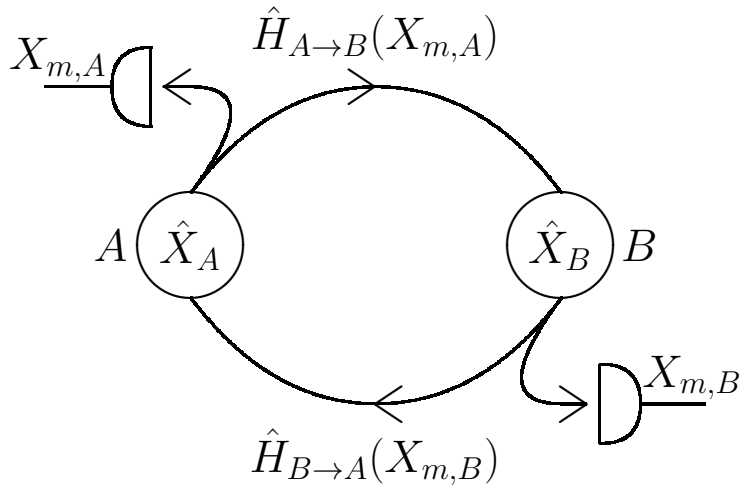
\begin{figure}
\begin{center}
\begin{picture}(280,200)
\thicklines
\put(30,90){\makebox(20,20){\Large $A$}}
\put(70,100){\circle{40}}
\put(60,90){\makebox(20,20){\Large $\hat{X}_A$}}

\put(230,90){\makebox(20,20){\Large $B$}}
\put(210,100){\circle{40}}
\put(200,90){\makebox(20,20){\Large $\hat{X}_B$}}

\bezier{200}(70,120)(100,160)(140,160)
\bezier{200}(140,160)(180,160)(210,120)
\put(140,160){\line(-3,2){10}}
\put(140,160){\line(-3,-2){10}}
\put(120,170){\makebox(40,20){\Large $\hat{H}_{A \to B}(X_{m,A})$}}
\bezier{200}(70,120)(100,160)(70,160)
\put(70,160){\line(-1,0){10}}
\put(60,160){\line(3,2){10}}
\put(60,160){\line(3,-2){10}}
\put(55,145){\line(0,1){30}}
\bezier{100}(55,145)(40,145)(40,160)
\bezier{100}(55,175)(40,175)(40,160)
\put(40,160){\line(-1,0){25}}
\put(0,160){\makebox(40,20){\Large $X_{m,A}$}}

\bezier{200}(70,80)(100,40)(140,40)
\bezier{200}(140,40)(180,40)(210,80)
\put(140,40){\line(3,2){10}}
\put(140,40){\line(3,-2){10}}
\put(120,10){\makebox(40,20){\Large $\hat{H}_{B \to A}(X_{m,B})$}}
\bezier{200}(210,40)(180,40)(210,80)
\put(210,40){\line(1,0){10}}
\put(220,40){\line(-3,2){10}}
\put(220,40){\line(-3,-2){10}}
\put(225,25){\line(0,1){30}}
\bezier{100}(225,25)(240,25)(240,40)
\bezier{100}(225,55)(240,55)(240,40)
\put(240,40){\line(1,0){25}}
\put(240,40){\makebox(40,20){\Large $X_{m,B}$}}
\end{picture}
\end{center}
\caption{\label{fig4}
Separation of quantum interaction dynamics by measurements of
the information emitted into the environment.
$\hat{X}_A$ and $\hat{X}_B$ can be measured with sufficient
precision so that the interaction dynamics can be identified
with the local Hamiltonians
$\hat{H}_{A \to B}(X_{m,A})$ and $\hat{H}_{B \to A}(X_{m,B})$
conditioned by the measurement outcomes $X_{m,A}$ and $X_{m,B}$.
}
\end{figure}

The special feature of an interaction realized entirely by quantum
feedback is that the conditional evolution of the two systems is fully
defined by the available classical information. In a direct
quantum interaction between two systems, this kind of information is
not necessarily available. However, it is possible that the interaction
of the systems with the environment makes such information available
even when the interaction is not implemented by feedback. In this case,
the interaction is too noisy to entangle the system, and the classical
information necessary to identify the local system dynamics is in
principle available in the local environments of the interacting
systems. Figure \ref{fig4} illustrates
this case: the measurement of $X_A$ in the emitted fields allows an
identification of the Hamiltonian $H_{A \to B}$ determining the evolution
of system $B$, and the measurement of $X_B$ in the emitted fields allows an
identification of the Hamiltonian $H_{B \to A}$ determining the evolution
of system $A$. The measurements in the environment can thus resolve the
entanglement between the two systems and the environment by projecting
the systems into a product state.

As figure \ref{fig4} suggests, the noise levels required to achieve this
identification of local dynamics are equal to the noise levels generated
by the corresponding quantum feedback based interaction. It is therefore
possible to give some quantitative limits beyond which the information
in the environment is definitely sufficient for an identification of
the local dynamics. This feedback-based analysis can then be used to
obtain quantitative results on the robustness of the entangling
capabilities of interaction Hamiltonians against quantum noise
\cite{Har03,Mon03,Ban04,Hof04a,Hof04b}.

If the system dynamics is described by
\begin{equation}
\label{eq:general}
\frac{d}{dt} \hat{\rho}_{AB} =
- \frac{i}{\hbar} [ \hat{H}_0, \hat{\rho}_{AB} ]
+ L(\hat{\rho}_{AB}),
\end{equation}
where $\hat{H}_0$ is the Hamiltonian describing the unitary part of
the dynamics and $L$ is the superoperator describing the non-unitary
part of the dynamics, the condition for separability of the dynamics
is given by a relation between the frequencies defining
$\hat{H}_0/\hbar$ and the rates defining $L$. Using the results of
section \ref{sec:int} above, it is possible to define the separability
limit for the case of a bilinear interaction in the presence of dephasing
between the eigenstates of the interaction Hamiltonian,
\begin{eqnarray}
\label{eq:model}
\hat{H}_0 &=& \hbar g_{AB} \hat{X}_A \hat{X}_B \hspace{0.5cm} \mbox{and}
\nonumber \\[0.2cm]
L(\hat{\rho}_{AB}) &=& \gamma_A \left(\hat{X}_A \hat{\rho}_{AB} \hat{X}_A
- \frac{1}{2} \hat{X}_A^2 \hat{\rho}_{AB} - \frac{1}{2} \hat{\rho}_{AB}
\hat{X}_A^2 \right)
\nonumber \\ &&
 + \gamma_B \left(\hat{X}_B \hat{\rho}_{AB} \hat{X}_B
- \frac{1}{2} \hat{X}_B^2 \hat{\rho}_{AB} - \frac{1}{2} \hat{\rho}_{AB}
\hat{X}_B^2 \right),
\end{eqnarray}
where the strength of the interaction is defined by the coupling frequency
$g_{AB}$ and the local decoherence rates of systems $A$ and $B$ are given by
$\gamma_A$ and by $\gamma_B$, respectively.
The separate variation of $\gamma_A$ and of $\gamma_B$ can be obtained
by rescaling the operators $\hat{X}_A$ and $\hat{X}_B$ in equation
(\ref{eq:fb3}) while leaving the product $\hat{X}_A \hat{X}_B$ unchanged.
This equation of motion for the density matrix can then be separated
into an effective local feedback scenario if (and only if)
\begin{equation}
g_{AB}^2 \leq \gamma_A \gamma_B.
\end{equation}
That this condition is in fact both necessary and sufficient is a 
consequence of the conservation of eigenstates of $\hat{X}_A$ and 
$\hat{X}_B$ in the noisy dynamics given by equations (\ref{eq:model}). 
The only possible local feedback model consistent with this conservation
of $\hat{X}_A$ and $\hat{X}_B$ is the model based on quantum non-demolition
measurements of $\hat{X}_A$ and $\hat{X}_B$, as given by equations 
(\ref{eq:qndAB}) to (\ref{eq:HeffAB}). Consequently, it is not possible
to construct any local feedback scenario for equations (\ref{eq:model}) 
if $g_{AB}^2 > \gamma_A \gamma_B$.

In the case of a sum of several bilinear interactions in the Hamiltonian,
sufficient conditions for separability can be obtained by simply combining
all individual separability criteria. Obtaining a necessary conditions
for separability is usually more difficult, since a
general superoperator $L$ can have infinitely many possible
decompositions \cite{Har03,Hof04a}. It is therefore generally unclear 
what selection of quantum measurements are optimal as a starting point
for the feedback model.

\section{Application to the two mode squeezing Hamiltonian}

\label{sec:squeeze}

It may now be instructive to consider a specific case of feedback
induced interactions in optical systems. The most simple example
is perhaps given by the dynamics of two resonant optical cavity modes
described by the annihilation operators
$\hat{a}_1=\hat{x}_1 + i \hat{y}_1$ and
$\hat{a}_2=\hat{x}_2 + i \hat{y}_2$,
respectively. If the attenuation rate of both cavities is $\kappa$,
the initial dynamics of the two cavity modes is simply given by
\begin{equation}
\label{eq:cavity}
\frac{d}{dt} \hat{\rho}_j = 2 \kappa
\left(\hat{a}_j \hat{\rho}_j \hat{a}_j^\dagger
- \frac{1}{2} \hat{a}_j^\dagger \hat{a}_j \hat{\rho}_j
- \frac{1}{2} \hat{\rho}_j \hat{a}_j^\dagger \hat{a}_j
\right),
\end{equation}
describing the emission of light from the cavities.
If the emitted light is detected by heterodyne detection, both quadrature
components can be measured with a resolution of
\begin{equation}
\frac{1}{\delta \! x^2} = \frac{1}{\delta \! y^2} = 4 \kappa \Delta t.
\end{equation}
Therefore, both components can be used to generate feedback, where the
cavity emission rate $\kappa$ corresponds to the decoherence rate
$\gamma$ of sections \ref{sec:qf} and \ref{sec:int}.

It is then possible to implement a two mode squeezing Hamiltonian by
using the following feedback Hamiltonians,
\begin{eqnarray}
\label{eq:Hcav}
\hat{H}_{x1\to x2} (x_{m,1}) = + 2 \hbar \kappa x_{m,1} \hat{x}_2,
\hspace{0.2cm} && \hspace{0.2cm}
\hat{H}_{x2\to x1} (x_{m,2}) = + 2 \hbar \kappa x_{m,2} \hat{x}_1,
\nonumber \\
\hat{H}_{y1\to y2} (y_{m,1}) = - 2 \hbar \kappa y_{m,1} \hat{y}_2,
\hspace{0.2cm} && \hspace{0.2cm}
\hat{H}_{y2\to y1} (y_{m,2}) = - 2 \hbar \kappa y_{m,2} \hat{y}_1.
\end{eqnarray}
If the total interaction dynamics of this feedback setup is written
as
\begin{equation}
\frac{d}{dt} \hat{\rho}_{12} =
- \frac{i}{\hbar} [\hat{H}_{\mbox{eff.}}, \hat{\rho}_{12}]
+ 2 D(\hat{\rho}_{12}),
\end{equation}
the effective two mode squeezing Hamiltonian is given by
\begin{equation}
\label{eq:squeeze}
\hat{H}_{\mbox{eff.}} = 2 \hbar \kappa
\left( \hat{x}_1\hat{x}_2 - \hat{y}_1\hat{y}_2 \right)
\; = \;
\hbar \kappa
\left( \hat{a}_1\hat{a}_2 + \hat{a}_1^\dagger\hat{a}_2^\dagger \right).
\end{equation}
In the absence of the decoherence represented by $D$, this two mode
squeezing would squeeze the noise in the two mode quadratures
$\hat{x}_1-\hat{x}_2$ and $\hat{y}_1 + \hat{y}_2$ to zero at an
exponential relaxation rate of $\kappa$, creating standard squeezed
state entanglement in the process. However, the decoherence effects
of the feedback add noise to this relaxation process according to
\begin{eqnarray}
\label{eq:fbnoise}
2 D(\hat{\rho}_{12}) &=& 3 \kappa
\left(\hat{a}_1 \hat{\rho}_{12} \hat{a}_1^\dagger
- \frac{1}{2} \hat{a}_1^\dagger \hat{a}_1 \hat{\rho}_{12}
- \frac{1}{2} \hat{\rho}_{12} \hat{a}_1^\dagger \hat{a}_1\right)
\nonumber \\ &&
+ \kappa \left(\hat{a}_1^\dagger \hat{\rho}_{12} \hat{a}_1
- \frac{1}{2} \hat{a}_1 \hat{a}_1^\dagger \hat{\rho}_{12}
- \frac{1}{2} \hat{\rho}_{12} \hat{a}_1\hat{a}_1^\dagger \right)
\nonumber \\ &&
 + 3 \kappa
\left(\hat{a}_2 \hat{\rho}_{12} \hat{a}_2^\dagger
- \frac{1}{2} \hat{a}_2^\dagger \hat{a}_2 \hat{\rho}_{12}
- \frac{1}{2} \hat{\rho}_{12} \hat{a}_2^\dagger \hat{a}_2\right)
\nonumber \\ &&
+ \kappa \left(\hat{a}_2^\dagger \hat{\rho}_{12} \hat{a}_2
- \frac{1}{2} \hat{a}_2 \hat{a}_2^\dagger \hat{\rho}_{12}
- \frac{1}{2} \hat{\rho}_{12} \hat{a}_2\hat{a}_2^\dagger \right).
\end{eqnarray}
This super operator still describes a relaxation of the
field components at a rate of $\kappa$, but the feedback has
doubled the rate at which quantum noise enters the cavities.
The relaxation dynamics of the squeezed field fluctuations is
therefore given by
\begin{eqnarray}
\frac{d}{dt} \langle (\hat{x}_1-\hat{x}_2)^2 \rangle
&=& -4 \kappa \langle (\hat{x}_1-\hat{x}_2)^2 \rangle + 2
\nonumber \\
\frac{d}{dt} \langle (\hat{y}_1+\hat{y}_2)^2 \rangle
&=& -4 \kappa \langle (\hat{y}_1+\hat{y}_2)^2 \rangle + 2,
\end{eqnarray}
and the stationary solutions are exactly equal to the vacuum
noise level of $1/2$.

As expected, the feedback interaction therefore cannot entangle
the cavity fields. However, it should be noted that even a
slight increase in the squeezing interaction will reduce the
noise level of both $\hat{x}_1-\hat{x}_2$ and $\hat{y}_1 + \hat{y}_2$
and lead to a violation of local uncertainties, indicating entanglement
\cite{Dua00,Hof03}. The feedback induced interaction described above
is therefore the strongest two mode squeezing interaction that can
be realized by local operations and classical communication only.
This result can be used to analyze the entangling capability of
a two mode squeezing interaction in the presence of noise according
to the procedure outlined in section \ref{sec:limits} above.
Specifically, a noisy two mode squeezing operation can be given
in the form defined by equation (\ref{eq:general}). The dynamics is
then described by the squeezing
Hamiltonian $\hat{H}_0$ and a decoherence operator $L$,
\begin{eqnarray}
\label{eq:noise}
\hat{H}_{0} &=& 2 \hbar g_{12}
\left( \hat{x}_1\hat{x}_2 - \hat{y}_1\hat{y}_2 \right)
\hspace{0.5cm} \mbox{and}
\nonumber \\[0.2cm]
L(\hat{\rho}_{12}) &=&
\gamma_-
\left(\hat{a}_1 \hat{\rho}_{12} \hat{a}_1^\dagger
- \frac{1}{2} \hat{a}_1^\dagger \hat{a}_1 \hat{\rho}_{12}
- \frac{1}{2} \hat{\rho}_{12} \hat{a}_1^\dagger \hat{a}_1\right)
\nonumber \\ &&
+ \gamma_+ \left(\hat{a}_1^\dagger \hat{\rho}_{12} \hat{a}_1
- \frac{1}{2} \hat{a}_1 \hat{a}_1^\dagger \hat{\rho}_{12}
- \frac{1}{2} \hat{\rho}_{12} \hat{a}_1\hat{a}_1^\dagger \right)
\nonumber \\ &&
+ \gamma_-
\left(\hat{a}_2 \hat{\rho}_{12} \hat{a}_2^\dagger
- \frac{1}{2} \hat{a}_2^\dagger \hat{a}_2 \hat{\rho}_{12}
- \frac{1}{2} \hat{\rho}_{12} \hat{a}_2^\dagger \hat{a}_2\right)
\nonumber \\ &&
+ \gamma_+ \left(\hat{a}_2^\dagger \hat{\rho}_{12} \hat{a}_2
- \frac{1}{2} \hat{a}_2 \hat{a}_2^\dagger \hat{\rho}_{12}
- \frac{1}{2} \hat{\rho}_{12} \hat{a}_2\hat{a}_2^\dagger \right).
\end{eqnarray}
Here, the squeezing rate is expressed by the coupling frequency $g_{12}$
and the decoherence rates are expressed by $\gamma_-$ for photon loss
and by $\gamma_+$ for photon gain ($\gamma_- > \gamma_+$ for stationary
solutions). Comparison with equations (\ref{eq:squeeze}) and
(\ref{eq:fbnoise}) then shows that the two mode squeezing Hamiltonian
could only originate from local measurements and feedback if
\begin{equation}
g_{12} \leq \gamma_+.
\end{equation}
This separability limit is consistent with the uncertainty limit
obtained from the steady state of the squeezing dynamics,
\begin{eqnarray}
\frac{d}{dt} \langle (\hat{x}_1-\hat{x}_2)^2 \rangle
&=& - (\gamma_- - \gamma_+ + 2 g_{12})
\langle (\hat{x}_1-\hat{x}_2)^2 \rangle
+ \frac{1}{2}(\gamma_- + \gamma_+)
\nonumber \\
\frac{d}{dt} \langle (\hat{y}_1+\hat{y}_2)^2 \rangle
&=& - (\gamma_- - \gamma_+ + 2 g_{12})
\langle (\hat{y}_1+\hat{y}_2)^2 \rangle
+ \frac{1}{2}(\gamma_- + \gamma_+).
\end{eqnarray}
Entanglement is obtained when the two quadrature components
drop below the standard quantum limit, a result that is
obtained in the steady state for $g_{12} > \gamma_+$.
The example of two mode squeezing thus illustrates how quantum
feedback scenarios can be used to identify the separability
limit of a specific entanglement generating Hamiltonian.

\section{Information inside and outside the system:
separability and its interpretation}

\label{sec:discuss}

The quantum feedback analysis of interactions between quantum
systems illustrates the correspondence between classical
interactions and quantum interactions by restoring the concept
of local forces to the quantum formalism. Instead of treating
the interaction process as an inseparable whole, it is now
possible to analyze the sequence of cause and effect in the
Hamiltonian dynamics. The difference between entanglement
generating dynamics and separable dynamics is then a quantitative
one, depending only on the noise levels of the systems.

Nevertheless, there remains one significant difference that
appears to introduce a qualitative element to the distinction
between quantum systems and classical systems. In the completely
classical interaction scenario shown in figure $\ref{fig1}$, we
would naturally identify the properties of the systems directly
with their measurement record in the environment. In the quantum
case, however, this identification of measurement record and
system property is not even legitimate in the separable case,
since the resolution of quantum measurements is necessarily
limited by quantum fluctuations. Consequently, it is not possible
to access the ``real'' system properties at all, and a description
of the systems in terms of the available information is all that
we can achieve.
In this sense, even local quantum systems do not have a classical
description. What then is the significance of the distinction
between separable and entangled systems?

Quantum feedback provides some insight into this fundamental
question by achieving a partial separation of information
and physical causality. In the case of the separable interaction
dynamics induced by quantum feedback, the information exchanged
between the systems is available as classical information
outside the systems. The information exchange between the
systems is therefore completely detached from the actual
system properties once the measurement has been performed.
This detachment of information and physical properties is
the decisive difference between the separable feedback
and the direct entanglement generating interaction. In the
pure Hamiltonian interaction, the exchange of quantum
information implies that the information about the exact
forces acting on each local system remains attached to the
physical properties of the systems. No information about the
interaction is available outside of the systems.
Separability therefore indicates that the information about
the interaction has detached itself from the quantum systems
and is now available as classical information in the
environment outside the systems, while entanglement generation
indicates that the quantum information shared between the
systems is exclusively confined inside the two interacting
systems.

\section{Conclusions}

\label{sec:concl}

As the analysis above has shown, the simulation of interaction
Hamiltonians by quantum feedback can provide
fundamental insights into the dynamics of information exchange
between interacting systems.
In particular, the dynamics expressed by an interaction Hamiltonian
can be interpreted in terms of a completely classical information
exchange if the decoherence rates are comparable to the frequencies
defining the strength of the interaction in the Hamiltonian.
The difference between a separable interaction that can be
represented by local operations and classical communication between
the systems and a genuine entanglement generating quantum interaction
can then be expressed quantitatively in terms of the decoherence
rates associated with the availability of information in the environment
outside the interacting systems. It may thus be possible to identify
the robustness of quantum interactions against various noise
effects by applying an appropriate quantum feedback model.
\ack
Part of this work has been supported by the JST-CREST project on
quantum information processing.


\end{document}